\documentclass[12pt]{article}
\usepackage{graphicx}
\usepackage{multirow}
\usepackage{booktabs}
\usepackage{braket}
\usepackage{subfigure}
\usepackage{epstopdf}
\usepackage{float}
\usepackage{amsmath,amssymb,amstext,amsfonts,array,hhline,tabularx,graphicx,epstopdf,}
\usepackage[numbers,sort&compress]{natbib}
\makeatletter

\renewcommand{\maketag@@@}[1]{\hbox{\m@th\normalsize\normalfont#1}}%
\UseRawInputEncoding

\makeatother

\begin{document}
	\parskip=3pt
	\parindent=18pt
	\baselineskip=20pt
	\setcounter{page}{1}

\title{Wave-Particle-Mixedness Relationships Based on $l_p$-norm Coherence
	\author{\ Wei-Ning Li $^{1}$ , Ming Fang $^{1}$ , Yuan-Hong Tao $^{2}$ \footnote{Corresponding author: Yuan-Hong Tao E-mail: taoyuanhong12@126.com}, Liu Sun $^{3}$ , Shao-Ming Fei $^{4}$ \\
		\footnotesize{1. College of Sciences, Yanbian University, Yanji 133002, China}\\
		\footnotesize{2. School of Science, Zhejiang University of Science and Technology, Hangzhou 310023, China}\\
		\footnotesize{3. College of Mathematics Science, Harbin Engineering University, Harbin 150001, China}\\
		\footnotesize{4. School of Mathematical Sciences, Capital Normal University, Beijing 100048, China  }\\
}}	

	\date{}
	\maketitle
	\date{}
	
	\textbf{\footnotesize{Abstract:}} \footnotesize{ We investigate the relationships among the wave property, particle property and mixedness of quantum states based on the \(l_p\)-norm coherence. By conforming that the \(l_p\)-norm coherence is an appropriate measure of wave property and introducing a measure of particle property based on the differences between the maximal \(l_2\)-norm coherence and the general \(l_2\)-norm coherence, we present tradeoff relationships among the wave, particle and mixedness of quantum states.
		For $1\leq p <2$,  we establish two kinds of tradeoffs of  the wave, particle and mixedness with respect to the upper and lower bounds of the \(l_p\)-norm coherence given by the \(l_2\)-norm coherence. These trade relations give rise to compressive understanding of the intrinsic connections among the  wave, particle and mixedness of quantum states, and cover some existing results as particular ones.
	}

	
	\begin{normalsize}
		
		\section{Introduction}

		\vspace{0.2cm}

Quantum coherence characterizes the superposition of quantum states and enables transformative technologies in quantum computing \cite{Barz2016,Shi2017,Rastegin2018,Ma2019a}, quantum key distribution \cite{Zhou2019,Ma2018} and quantum teleportation \cite{Bennett1993}. The framework for quantifying the quantum coherence was systematically established by Baumgratz et al. in 2014 \cite{Baumgratz2014}, which spurred the advances on various coherence quantifiers \cite{Baumgratz2014,Chen2018,Yu2016,Shao2015,Liu2017,Rastegin2016,Shao2017,Yu2017,Streltsov2015,Xiong2018,Xiong2018}.

Due to the decoherence \cite{Zeh1970}, the coherence may diminish and increase the mixedness  of the quantum system, revealing a trade-off between these two crucial quantities. Given such inherent trade-off, the challenge of formulating a comprehensive relationship between coherence and mixedness has emerged as a critical area of research. In 2015, Uttam Singh et al. \cite{Singh2015} initiated the exploration by studying the trade-off between \(l_1\)-norm coherence and the normalized linear entropy mixedness. Since then, numerous studies \cite{Shao2015,Shao2017,Song2020,Che2023,Sun2023_lp,Zhang2024,Peters2004} have delved into the trade-offs involving various coherence measures such as \(l_2\)-norm coherence, \(l_p\)-norm coherence, skew information coherence and R{\'e}nyi-\(\alpha\) entropy coherence, as well as different mixedness measures.

The interplay between quantum coherence and mixedness offers an essential vantage point for deciphering the dynamics of quantum systems. This trade-off not only mirrors how quantum resources evolve amidst environmental noise but also intricately ties into the fundamental wave-particle duality inherent to quantum systems. The wave-particle duality, a fundamental concept embodying Bohr's principle of complementarity, is commonly quantified through the path predictability and the interference visibility. In 2022, Fu et al. \cite{Fu2022} formulated an equation concerning wave-particle-mixedness in terms of variance.
Research on wave-particle duality has unearthed the intricate connections and interactions among the predictability, distinguishability, visibility  \cite{Wootters1979,Greenberger1988,Englert1996,Durr2001,Peng2005,Englert2008},  coherence, mixedness, entanglement and other correlations \cite{Tsui2024,Yang2024,Bai2025,Machado2024,Ding2025,Sun2023,Luo2017,Luo2018,Hu2018,Bu2018,Bagan2018,Roy2019,Sun2021,Basso2021}.

Recently, Tsui et al. \cite{Tsui2024} proposed a generalized wave-particle-mixedness triality by using symmetric concave functions. Yang et al. \cite{Yang2024} advanced this research by establishing another generalized triality through operator concave functions, in which  the complementarity associated with the $l_p$-norm coherence is excluded, and the measure of mixedness is basis-dependent. Building upon Tsui's work, Bai et al. \cite{Bai2025} extended the concept to a more general triality applicable to any valid coherence measures, although  the measure of mixedness  is still basis-dependent.
In 2024, P. Machado et al. \cite{Machado2024} adopted the $l_1$-norm coherence and introduced an additional variable. They successfully derived an equality satisfied by the wave, particle, mixedness and the additional variable.

This paper primarily focuses on establishing wave-particle-mixedness relationships grounded in the $l_p$-norm ($1\leq p\leq 2$) coherence. In Section II, we recall some related concepts and conform that the $l_p$-norm coherence is a bona fide measure of wave property. Section III delves into the derivation of the wave-particle-mixedness triality under the $l_2$-norm coherence, which includes the existing results as special ones. In Section IV, we introduce two new variables by leveraging the magnitude relationship between the \(l_p\)-norm and \(l_2\)-norm, and establish two kinds of tradeoff relations among the introduced variable, wave, particle and mixedness. The physical implications of the new variables are illustrated in terms of the distribution characteristics of quantum coherence. Finally, Section VI summarizes the key findings, highlights the contributions of the proposed relationships.
	
\section{Preliminaries}

Consider a general $d$-dimensional quantum state,
\begin{equation}\label{rho}
\rho=\sum_{i,j=1}^{d}\rho_{ij}|i\rangle\langle j|,
\end{equation}
where $\{|i\rangle,i=1,2,...,d\}$ is the fixed computational basis and $\rho_{ij}=\langle i| \rho|j\rangle$. For a quantum state $\rho$ described within a set of paths $\{|i\rangle,i = 1,2,\ldots,d\}$, the measure $V(\rho)$ of wave property should adhere to the following well-founded conditions \cite{Durr2001}:
	
(a1) $V(\rho)$ reaches its global minimum if the state $\rho$ is diagonal with respect to the fixed basis;
	
(a2) $V(\rho)$ reaches its global maximum if the state $\rho$ is pure and a uniform superposition in the fixed basis;
	
(a3) $V(\rho)$ is invariant under permutations of the diagonal elements of $\rho$;
	
(a4) $V(\rho)$ is convex on $\rho$.
	
Similarly, the measure $P(\rho)$ of particle property should satisfy the following conditions:

(b1) $P(\rho)$ reaches its global maximum if $\left|\rho_{ii}\right|=1$ for some $i$;
	
(b2) $P(\rho)$ reaches its global minimum if $\left|\rho_{ii}\right|=\frac{1}{d}$ for all $i$;
	
(b3) $P(\rho)$ is invariant under permutations of the diagonal elements of $\rho$;
	
(b4) $P(\rho)$ is convex on $\rho$.

The $l_p$-norm coherence $C_{l_{p}}(\rho)$ of a quantum state $\rho$ is defined by \cite{Baumgratz2014,Song2020},
\begin{equation}\label{l_p}
C_{l_{p}}(\rho)=\left(\sum_{i\neq j}|\rho_{ij}|^p\right)^{\frac{1}{p}}, \quad p \geq 1.
\end{equation}
Specifically, when $p = 1$, the $l_1$-norm coherence $C_{l_1}(\rho)$ satisfies all the criteria of a proper coherence measure, including the strong monotonicity under incoherent operations. However, for $p > 1$,  $C_{l_p}(\rho)$ only meets the requirements of a coherence monotone, but not the strong monotonicity. The $l_p$-norm coherence $(1 \le p\le 2)$ gives rise to a collection of suitable measures of wave property.

\textbf{Lemma 1.}
The $l_p $-norm $(1 \le p\le 2)$ coherence is a collection of suitable wave measures.

A clear and concise proof of Lemma 1 is provided in Appendix A, see also \cite{Ding2025} for an alternative proof. It is worth noting that the special case \( p = 1 \) has been addressed in the relevant literature \cite{Bera2015}.

The normalized linear entropy mixedness $M_l(\rho)$ of a state $\rho$ is given by \cite{Peters2004}
\begin{equation}\label{M_l}
M_l(\rho)=\frac{d}{d-1}\left(1-\text{tr}\rho^2\right).
\end{equation}
We study next the trade-offs between the wave-particle duality associated with the \( l_p \)-norm coherence and the mixedness, serving as a key indicator that encapsulates the intrinsic disorder level and uncertainty of the quantum system. We investigate the problem for the  case of \( p = 2 \) in Section III and the case of \( 1 \le p < 2 \) in Section IV.

\section{Wave-particle-mixedness triality based on $l_2$-norm coherence}\label{sec:section3}

Based on the \( l_2 \)-norm coherence, we have the following conclusion.

\textbf{Lemma 2.} The $P(\rho)$ defined by
$P(\rho)  = \sqrt{\frac{d}{d - 1}\left(C_{l_{2}}^{\max}\left(\rho\right)\right)^{2} - C_{l_{2}}^{2}\left(\rho\right)}$
is a bona fide measure of particle property,  where $C_{l_{2}}^{\mathrm{max}}\left(\rho\right)=\left(\text{tr}\rho^{2}-\frac{1}{d}\right)^{\frac{1}{2}}$ denotes the maximal coherence calculated across all reference bases \cite{Zhang2024}.

\textbf{Proof.}
By straightforward calculation we have
\[
\begin{aligned}
	\left(C_{l_{2}}^{\max}\left(\rho\right)\right)^{2} - C_{l_{2}}^{2}\left(\rho\right) & = \text{tr} \rho^{2} - \frac{1}{d} - \sum_{i\neq j}\left|\rho_{ij}\right|^{2} \\
	& = \sum_{i,j}\left|\rho_{ij}\right|^{2} - \frac{1}{d} - \sum_{i\neq j}\left|\rho_{ij}\right|^{2} \\
	& = \sum_{i}\left|\rho_{ii}\right|^{2} - \frac{1}{d} \\
	& = \frac{1}{2}P_{JB}^2(\rho),
\end{aligned}
\]
where ${P_{JB}} = \sqrt{2\left(\sum_{i}^{d}\left|\rho_{ii}\right|^{2} - \frac{1}{d}\right)}$ is just the measure of particle property presented in \cite{Jakob2007}.

We next establish the tradeoff among the $l_2$-norm coherence, particle measure $P(\rho)$ and the normalized linear entropy mixedness $M_l(\rho)$.

\textbf{Theorem 1.}
For any $d$-dimensional quantum system, we have following tradeoff  of wave-particle-mixedness associated with the $l_{2}$-norm coherence,
	\begin{equation}\label{l2}
\frac{d}{d - 1}C_{l_{2}}^{2}\left(\rho\right) + M_{l}\left(\rho\right) + P^{2}(\rho) = 1.
\end{equation}

\textbf{Proof.}​ It has been shown that \cite{Zhang2024}
\begin{equation}\label{max}
\frac{d}{d - 1}\left[C_{l_{2}}^{\mathrm{max}}\left(\rho\right)\right]^{2}+M_{l}\left(\rho\right)=1.
\end{equation}
From \eqref{max} it is straightforward to observe that
\begin{equation*}
\frac{d}{d - 1}C_{l_{2}}^{2}\left(\rho\right) + M_{l}\left(\rho\right) + \frac{d}{d - 1}\left[\left(C_{l_{2}}^{\max}\left(\rho\right)\right)^{2} - C_{l_{2}}^{2}\left(\rho\right)\right] = 1,
\end{equation*}
which proves the equality \eqref{l2}.

Concerning the wave-particle-mixedness triality, D{\"u}rr \cite{Durr2001} derived the following relationship in terms of the interference patterns and probability distributions from a \(d\)-slit interferometer,
\begin{equation}\label{D}
\text{tr}\rho^{2}=\frac{1}{d}+\frac{d - 1}{d}(P_{\mathrm{D}}^{2}(\rho|\Pi)+V_{\mathrm{D}}^{2}(\rho|\Pi)),
\end{equation}
where \(V_{D}\left(\rho|\Pi\right)=\sqrt{\frac{d}{d-1}\sum_{i\neq j}^{d}\left|\rho_{ij}\right|^{2}}\) quantifies the visibility and \(P_{D}\left(\rho|\Pi\right)=\sqrt{\frac{d}{d-1}\sum_{i=1}^{d}\left(\left|\rho_{ii}\right|-\frac{1}{d}\right)^{2}}\) represents the predictability. Jakob and Bergou \cite{Jakob2007} presented the following relationship,
\begin{equation}\label{JB}
{P_{JB}}^{2}+{V_{JB}}^{2}=2\left(\text{tr} \rho^{2}-\frac{1}{d}\right),
\end{equation}
where the visibility ${V_{JB}}=\sqrt{2\sum_{i\neq j}^{d}\left|\rho_{ij}\right|^{2}}$ quantifies the wave property and the predictability ${P_{JB}}=\sqrt{2\left(\sum_{i=1}^{d}\left|\rho_{ii}\right|^{2}-\frac{1}{d}\right)}$ measures the particle property.
Fu and Luo \cite{Fu2022} developed a new framework integrating quantum uncertainty, coherence and path information,
\begin{equation}\label{fl}
 P_{L}\left(\rho|\Pi\right)+V_{L}\left(\rho|\Pi\right)+\frac{d-1}{d}M_{l}\left(\rho\right)=1,
\end{equation}
where ${V_L}(\rho|\Pi)=\sum_{i\neq j}^{d}\left|\rho_{ij}\right|^{2}$ and ${P_L}(\rho|\Pi)=\sum_{i=1}^{d}\left|\rho_{ii}\right|^{2}$ quantify the wave and particle properties, respectively. The visibility is equivalent to the $l_2$-norm coherence (up to a constant factor). We have the following remarks.

\textbf{Remark 1.} An inequality satisfied by the $l_2$-norm coherence and the mixedness has been derived in \cite{Song2020,Che2023},
\begin{equation}\label{inequal-l2}
\frac{d}{d-1}C_{l_{2}}^{2}\left(\rho\right) + M_{l}\left(\rho\right) \leq 1.
\end{equation}
It is observed that from our Theorem the above inequality becomes a strict equality by adding the term $P^{2}(\rho)$
given by the measure of particle property, namely, we adopt the coherence difference as a measure of particle property to establish the wave-particle-mixedness triality. This framework yields mathematical relations analogous to those derived from physically distinct methodologies, as explicitly summarized in TABLE \ref{tab:triality}.
\begin{table*}[htbp]
	\centering	
	\caption{Comparisons among different triality relations}
	\renewcommand{\arraystretch}{2}
	\resizebox{\textwidth}{!}{
	\begin{tabular}{lllll}
		\hline  %
		& D{\"u}rr \cite{Durr2001} & Jacob and Bergou \cite{Jakob2007} & Fu and Luo \cite{Fu2022} & Our result \\
		\hline  %
		Wave & $V_D(\rho|\Pi) = \sqrt{\frac{d}{d-1}\sum_{i\neq j}|\rho_{ij}|^2}$ & $V_{JB}(\rho|\Pi) = \sqrt{2\sum_{i\neq j}|\rho_{ij}|^2}$ & $V_L(\rho|\Pi) = \sum_{i\neq j}|\rho_{ij}|^2$ & $C_{l_{2}}(\rho)=\big({\sum_{i\neq j}|\rho_{ij}|^2}\big )^{\frac{1}{2}}$\\
		
		Particle & $P_D(\rho|\Pi) = \sqrt{\frac{d}{d-1}\sum_{i=1}^{d}\left( |\rho_{ii}| - \frac{1}{d} \right)^2}$ & $P_{JB}(\rho|\Pi) = \sqrt{2\left( \sum_{i=1}^{d}|\rho_{ii}|^2 - \frac{1}{d} \right)}$ & $P_L(\rho|\Pi) = \sum_{i=1}^{n}|\rho_{ii}|^2$ & $P(\rho)=\sqrt{\frac{d}{d - 1}\left(\sum_{i = 1}^{d}\left|\rho_{ii}\right|^{2}-\frac{1}{d}\right)}$ \\
		
		Mixedness & $1 - \mathrm{tr}\rho^2 + \frac{1}{d}$ & Not mentioned & $1-tr\rho^2$ & $\frac d{d-1}(1-tr\rho^2)$ \\
		
		Triality & Equation \eqref{D} & Equation \eqref{JB} & Equation \eqref{fl} & Equation \eqref{l2} \\
		\hline  %
	\end{tabular}
}
	\label{tab:triality}
\end{table*}

\textbf{Remark 2.} In \cite{Ding2025} the authors established tradeoffs between wave-particle duality and quantum entanglement. In particular, the tradeoffs give rise to  our equation \eqref{l2} when the \(l_2\)-norm coherence is taken into account, indicating that the mixedness of a single state also characterizes the entanglement with respect to a purified bipartite state.
	
\section{Wave-particle-mixedness tradeoffs based on $l_p$-norm coherence ($1\le p<2$)}\label{sec:section5}

We present two types of wave-particle-mixedness trade-offs based on the $l_p$-norm coherence ($1\le p<2$).

\subsection{Wave-particle-mixedness-$X$ trade-off}

We first present the following lemma, see the detailed derivation in Appendix B.

\textbf{Lemma 3.}
The $l_p$-norm coherence ($1\le p<2$) and $l_2$-norm coherence satisfy the following inequality,
\begin{equation}\label{ineq lp}
C_{l_{2}}^{2}\left(\rho\right) \ge \left[d\left(d - 1\right)\right]^{\frac{p - 2}{p}}C_{l_{p}}^{2}\left(\rho\right),~  1\leq p<2.
\end{equation}

Denote
\begin{equation}\label{12}
X = C_{l_{2}}^{2}\left(\rho\right)-\left[d\left(d - 1\right)\right]^{\frac{p - 2}{p}}C_{l_{p}}^{2}\left(\rho\right),
\end{equation}
to be the difference between the left and right-hand sides of \eqref{ineq lp}.
From Lemma 2 it is direct to see that $X\geq 0$. By using Theorem 1, it is direct to verify the following results on the wave-particle-mixedness trade-off relation:

\textbf{Theorem 2.}
For any $d$-dimensional quantum system, the following tradeoffs hold among the $l_p$-norm coherence $(1\leq p<2)$, measure of particle property and normalized linear entropy mixedness,
\begin{equation}\label{lp - 4}
\left(\frac{d^{p - 1}}{d -1}\right)^{\frac{2}{p}}C_{l_{p}}^{2}\left(\rho\right)+M_{l}\left(\rho\right)+P^{2}\left(\rho\right)+\frac{d}{d-1}X = 1.
\end{equation}

\textbf{Remark 3.} It is easily seen that the following main results of Singh et al. (2015) \cite{Singh2015}, Che et al. (2023) \cite{Che2023} and Sun et al. (2023) \cite{Sun2023},
\begin{equation}\label{lp and Ml}
\left(\frac{d^{p-1}}{d-1}\right)^{\frac{2}{p}}C_{l_{p}}^{2}\left(\rho\right)+M_{l}\left(\rho\right)\leq1, \quad 1\leq p\leq 2,
\end{equation}
are direct corollaries of our Theorem 1 and Theorem 2.
 
From a geometric perspective, the quantity $X$ in Theorem 2 represents the difference between \(C_{l_2}^2(\rho)\) and \(C_{l_p}^2(\rho)\). When all the non-zero off-diagonal elements of $\rho$ are equal, the value of $X$ depends on both the number of off-diagonal elements (characterizing coherence distribution in the quantum states) and the magnitude of
these elements (reflecting the variation in the intensity of 
coherence within quantum states).

From a physical standpoint, the variable \(X\) plays a pivotal role in quantifying the partial coherence between quantum states, especially when \(\left|\rho_{ij}\right|\) is used to gauge the ``partial coherence" between basis states \(\left|i\right\rangle\) and \(\left|j\right\rangle\). Consider a scenario where coherence is entirely concentrated on a specific basis pair, say \(\{|i_0\rangle,|j_0\rangle\}\). In this case, \(X=\left[2 - 2^{\frac{2}{p}}\left[d\left(d - 1\right)\right]^{\frac{p - 2}{p}}\right]\left|\rho_{i_0j_0}\right|^{2}\) reveals a direct proportionality between \(X\) and the magnitude of \(\left|\rho_{i_0j_0}\right|^{2}\). When a quantum state has only one pair of non-zero off-diagonal elements, then \(C_{l_p}^2(\rho)\) is a constant multiple of \(X\). This characteristic allows \(X\) to act as a sensitive measure for coherence, effectively capturing the wave behavior of the quantum state. In the context of the maximally coherent mixed state (MCMS), $\rho_{\mathrm{MCMS}}=a\left|\phi_{d}\right\rangle\left\langle\phi_{d}\right|+\frac{1-a}{d}I_{d}$,
with $\left|\phi_{d}\right\rangle=\frac{1}{\sqrt{d}}\sum_{i = 1}^{d}\left|i\right\rangle$ and $a\in (0,1)$, a state that the cumulative ``partial coherence" across all basis pairs reaches its peak, \(X\) attains its minimum value. This demonstrates an inverse relationship: as the overall coherence of the system maximizes, the value of \(X\) minimizes, providing a clear physical benchmark for understanding the coherence properties within quantum states.

To elucidate vividly the equality \eqref{lp - 4} in Theorem 2, let us consider a distinctive tetrahedron model shown in FIG. \ref{fig:four}, where each face, line segment and vertex correspond to a particular subset of states within \eqref{lp - 4}.

\begin{figure}[htb]
	\centering
	\includegraphics[width=1\linewidth]{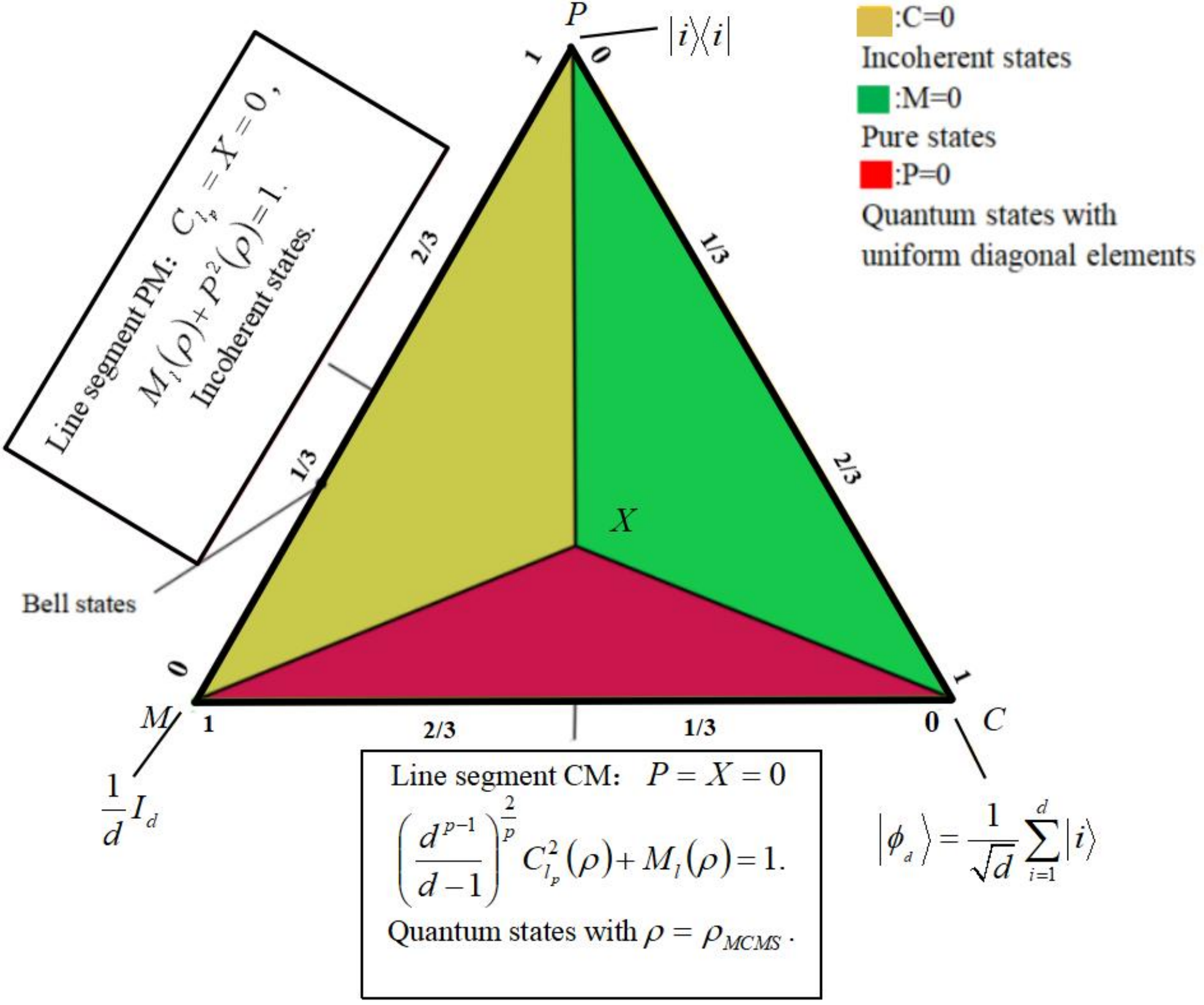}
	\caption{The tetrahedron representation equality \eqref{lp - 4}.}\label{fig:four}
\end{figure}

{\boldmath$C_{l_{p}}(\rho)=0$} when \(\rho\) is fully incoherent, namely, \(|\rho_{ij}| = 0\) for all \(i\neq j\). As a direct consequence of this property, \(X\) also vanishes. In this case \eqref{lp - 4} simplifies to
$M_{l}\left(\rho\right)+P^{2}\left(\rho\right)=1$. Geometrically, this particular scenario corresponds to the line segment \(MP\) illustrated in FIG. \ref{fig:four}, providing a visual representation of the relationship under the condition of vanishing coherence.

{\boldmath$X=0$} when \(\rho\) adheres to the condition \(|\rho_{ij}|=|\rho_{kl}|\) for all \(i\neq j\) and \(k\neq l\). To distinguish this scenario from the case of \(C_{l_p}(\rho)=0\), we exclude the incoherent states from our analysis.
Under these specific conditions, \eqref{lp - 4} reduces to
   \begin{equation}\label{X=0}
   \left(\frac{d^{p - 1}}{d - 1}\right)^{\frac{2}{p}}C_{l_{p}}^{2}\left(\rho\right)+M_{l}\left(\rho\right)+P^{2}\left(\rho\right)=1,
   \end{equation}
as geometrically represented by the bottom face \(CPM\) of the tetrahedron shown in FIG. \ref{fig:four}.
For a \(d\)-dimensional quantum state with \(X = 0\), the maximum value of the \(l_p\)-norm coherence, \(C_{l_{p}}\left(\rho\right)_{\mathrm{max}}=\left(\frac{d - 1}{d^{p - 1}}\right)^{\frac{1}{p}}\), is obtained when \(\rho=|\varphi\rangle\langle\varphi|\), where \(|\varphi\rangle=\frac{1}{\sqrt{d}}\sum_{i}|i\rangle\). For this particular state, \(M_{l}\) and \(P\) satisfy \(M_{l}(|\varphi\rangle\langle\varphi|)+P(|\varphi\rangle\langle\varphi|)^{2}=0\).

{\boldmath\(M_{l}(\rho)=0\)} when \(\rho\) is a pure state, namely, \(M_{l}(\rho)\) vanishes. In this case, \eqref{lp - 4} reduces to
\[
\left(\frac{d^{p - 1}}{d - 1}\right)^{\frac{2}{p}}C_{l_{p}}^{2}\left(\rho\right)+P^{2}\left(\rho\right)+\frac{d - 1}{d}X = 1,
\]
see the green-colored face \(CPX\) of the tetrahedron model of FIG. \ref{fig:four}. In this situation, \(P(\rho)^{2}=1\), while \(X = C_{l_p}^{2}(\rho)=0\). In the tetrahedron depicted in FIG. \ref{fig:four}, this specific condition is represented by the vertex \(P\), signifying a particular extreme case.

{\boldmath$P(\rho)=0$} when the diagonal elements of the quantum state \(\rho\) are all equal to \(\frac{1}{d}\). In this case, the measure \(P(\rho)\) of the particle property becomes zero, and \eqref{lp - 4} simplifies to
\[
\left(\frac{d^{p - 1}}{d - 1}\right)^{\frac{2}{p}}C_{l_{p}}^{2}\left(\rho\right)+M_{l}\left(\rho\right)+\frac{d - 1}{d}X = 1.
\]
This situation corresponds to the red-colored face \(CMX\) of the tetrahedron illustrated in FIG. \ref{fig:four}. This visual representation within the tetrahedron model offers an intuitive way to understand the relationships among the measures \(C_{l_{p}}(\rho)\), \(M_{l}(\rho)\) and \(X\) when \(P(\rho)\) vanishes, highlighting the unique characteristics of the quantum state under this specific condition.

Furthermore,  for completely mixed state (\(\rho = \frac{I}{d}\)), we have
\(X = 0\), \(C_{l_p}(\rho) = 0\) and \(M_l(\rho) = 1\), which corresponds to the vertex M  in FIG. \ref{fig:four}.
For maximally coherent state \(\rho = |\varphi\rangle\langle\varphi|\) with 
\(|\varphi\rangle = \frac{1}{\sqrt{d}}\sum_i |i\rangle\)), \(C_{l_p}(\rho)\) achieves its maximum \(\left(\frac{d-1}{d^{p-1}}\right)^{\frac{1}{p}}\), a case corresponding to the vertex C in FIG. \ref{fig:four}.
For the state with equal off-diagonal elements, (\(|\rho_{ij}| = |\rho_{kl}|\neq0, \ i\neq j, \ k\neq l\)), \eqref{lp - 4} reduces to \(\frac{1}{(d-1)^2}C_{l_p}^2(\rho) + M_l(\rho) = 1\), in correspondence to the line segment CM in FIG. \ref{fig:four}.

Due to the constraints imposed on the variables $X$ and the $l_p$-norm coherence $C_{l_p}(\rho)$, the simplified trade-off equations corresponding to the line segments $XP$ and $XM$ cannot be achieved in practical scenarios. The line segments $CX$ and $CP$ are uniquely associated with the maximally coherent state represented by the vertex $C$. The maximally coherent state is a pure quantum state with all off-diagonal elements being equal. As such, these two line segments do not represent a variety of physical states. The foregoing analysis underscores the mutually restrictive nature within the wave-particle-mixedness trade-off framework, highlighting how changes in one aspect of the quantum state necessarily impact the others.

\subsection{Wave-particle-mixedness-$Y$ trade-off}	

By leveraging the H{\"o}lder's inequality, we obtain the lower bound of \(l_p\)-norm coherence given by the \(l_2\)-norm coherence for $p\geq 1$, see the derivation in Appendix C,
\begin{equation}\label{lple}
\frac{1}{2}\left[d(d - 1)\right]^{\frac{2(p - 1)}{p}} C_{l_p}^2\left(\rho\right)\ge C_{l_2}^2\left(\rho\right).
\end{equation}
Based on above inequality, we define
\begin{equation}\label{19}
Y=\frac{1}{2}\left[d(d - 1)\right]^{\frac{2(p - 1)}{p}} C_{l_p}^2\left(\rho\right)-C_{l_2}^2\left(\rho\right)
\end{equation}
to be the difference between the left and right-hand sides of \eqref{lple}.

From \eqref{l2} in Theorem 1 it is straightforward to verify the following conclusion.

\textbf{Theorem 3.}
For any quantum state in a $d$-dimensional Hilbert space, the following equality holds,
\begin{equation}\label{lp-4-1}
\frac{d^{\frac{3p-2}{p}}(d-1)^{\frac{p-2}{p}}}{2}C_{l_{p}}^{2}\left(\rho\right)+M_{l}\left(\rho\right)+P^{2}\left(\rho\right) = 1+\frac{d}{d-1}Y.
\end{equation}

{\bf Remark 4.} When $p=1$, our relation \eqref{lp-4-1} turns to
\begin{equation}\label{l1Y}
	\frac{d}{2(d-1)}C_{l_{1}}^{2}\left(\rho\right)+M_{l}\left(\rho\right)+P^{2}\left(\rho\right) = 1+\frac{d}{d-1}Y,
\end{equation}
where $Y=\frac{1}{2}C_{l_1}^2\left(\rho\right) - C_{l_2}^2\left(\rho\right)$.
In \cite{Machado2024} P. Machado presented the following wave-particle-mixedness complementary relation,
\begin{equation}\label{PM}
	\frac{d}{2(d-1)}C_{l_1}^{2}(\rho)+M_l(\rho)+P^{2}(\rho)=1+\frac{(d-2)(d+1)}{2}T,
\end{equation}
where $T=\frac{4d}{(d-2)(d^{2}-1)}\sum_{j>i}\sum_{(i,j)\neq(k,l)}|\rho_{ij}||\rho_{kl}| $.
It is easy to prove that $\frac{(d-2)(d+1)}{2}T=\frac{d}{d-1}(C_{l_{1}}^{2}(\rho)-2C_{l_{2}}^{2}(\rho))=\frac{d}{d-1}Y$. Therefore, \eqref{PM} and \eqref{l1Y} are the same. Namely,  \eqref{PM} is a particular case of our Theorem with $p=1$.

Let's now delve into both the mathematical and physical interpretations of the quantity \(Y\) for general case \((1\leq p < 2)\). From equation \eqref{ineq lp}, $Y$ is bounded by the following upper and lower limits,
\begin{equation}
0\le Y\le \frac{(d - 2)(d + 1)}{2d(d - 1)}\left[d(d - 1)\right]^{\frac{2(p - 1)}{p}}C_{l_p}^2\left(\rho\right).
\end{equation}
When \(Y\) attains zero, the corresponding quantum states conform to the subsequent criteria:
their density matrices contain no more than one pair of non-zero off-diagonal elements. To clearly distinguish this case from the scenario where \(C_{l_p}(\rho)=0\), we temporarily exclude the consideration of incoherent states. Under these circumstances, we obtain
\begin{equation}\label{Y=0}
	\frac{d^{\frac{3p-2}{p}}(d - 1)^{\frac{p - 2}{p}}}{2}C_{l_{p}}^{2}\left(\rho\right)+M_{l}\left(\rho\right)+P^{2}\left(\rho\right)=1.
\end{equation}

When \(Y\) reaches its upper bound, if and only if \(|\rho_{ij}| = |\rho_{kl}|\) for all \(i\neq j\) and \(k\neq l\), which can be the case for coherent states when \(d>2\).
Under these conditions, the exact tradeoff of wave-particle-mixedness is given by
\begin{equation}\label{Y=0 le}
	\left(\frac{d^{p - 1}}{d - 1}\right)^{\frac{2}{p}}C_{l_{p}}^{2}\left(\rho\right)+M_{l}\left(\rho\right)+P^{2}\left(\rho\right)=1.
\end{equation}
This is the same as \eqref{X=0} (the case of \(X = 0\)) and as \eqref{Y=0} when \(d = 2\).

\subsection{Discussions on the relations between the wave-particle-mixedness-$X$  and $Y$ trade-offs}

Concerning the tradeoffs \eqref{12} and \eqref{19},  a simple calculation yields the following relation between  $X$  and  $Y$,
\begin{equation}\label{X+Y}
	X + Y = \left( \frac{1}{2} \left[ d(d-1) \right]^{\frac{2p - 2}{p}} - \left[ d(d-1) \right]^{\frac{p - 2}{p}} \right) C_{l_p}^2(\rho).
\end{equation}

Let us conduct a comparative analysis specifically for the case of \(p = 1\).
In this case \eqref{lp - 4} transforms into
\begin{equation}\label{l1}
\frac{1}{\left(d-1\right)^{2}}C_{l_{1}}^{2}\left(\rho\right)+M_{l}\left(\rho\right)+P^{2}\left(\rho\right)+\frac{d}{d-1}X=1,
\end{equation}
with $X=C_{l_{2}}^{2}\left(\rho\right)-\frac{1}{d\left(d-1\right)}C_{l_{1}}^{2}\left(\rho\right)$.
And \eqref{lp-4-1} turns into \eqref{l1Y} with $Y=\frac{1}{2}C_{l_1}^2\left(\rho\right) - C_{l_2}^2\left(\rho\right)$.
When the quantum state is uniformly distributed (prominent wave behavior), \(Y\) is maximized and \(X\) is minimized, both above equations reduce to the same triality relation  \eqref{X=0}. If coherence concentrates only on \(\{|i_0\rangle,|j_0\rangle\}\) (\(|\rho_{i_0j_0}|\neq 0\) and other off-diagonal elements are zero), \(Y\) is minimized while \(X\) remains non-zero. 

As a means to contextualize the link between \(X\) and \(Y\), we specifically consider Eq. \eqref{X+Y}, which emerges as a special instance of the formula established earlier,
\begin{equation}\label{XY}
X + Y=\frac{(d - 2)(d + 1)}{2d(d - 1)}C_{l_1}^2(\rho),
\end{equation}
which reveals their complementarity under \(l_1\)-norm coherence. Assuming that the off-diagonal elements take the same value, \(X\), \(Y\), and \(C_{l_1}^2(\rho)\) all depend on off-diagonal element distribution: for fixed number of non-zero off-diagonals, \(X\), \(Y\) and \(C_{l_1}^2(\rho)\) are monotonically increasing with respect to the modulus of the non-zero off-diagonal elements; for fixed non-zero off-diagonal elements: \(Y\) and \(C_{l_1}^2(\rho)\) increase monotonically with respect to the number of non-zero off-diagonal elements, while \(X\) rises and then falls, see Fig. \ref{Fig:XY} for \(d = 3\),
where (a) shows how \(X\), \(Y\) and \(\frac{1}{3}C_{l_1}^2\) vary with the number \(n\) of \(|\rho_{ij}| \neq 0\) (\(i < j\)) with fixed \(|\rho_{ij}|=\frac{1}{3}\), (b) shows how \(X\), \(Y\) and \(\frac{1}{3}C_{l_1}^2\) vary with the value \(m\) of \(|\rho_{ij}|\) for fixed number of non-zero elements \(n = 2\).
\begin{figure}[htb]
	\includegraphics[width=0.8\textwidth]{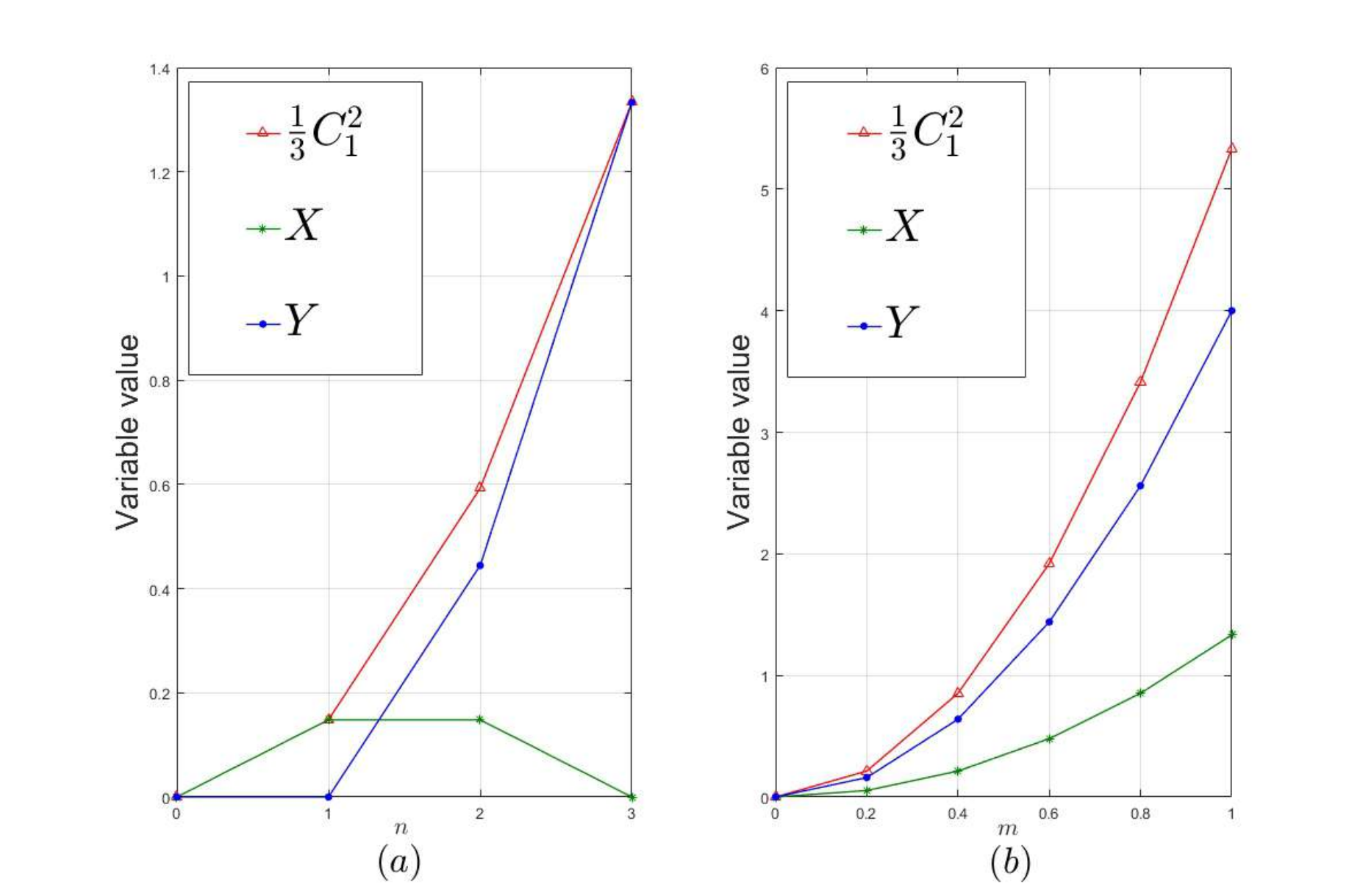}
	\caption{The behavior of \(X\), \(Y\) and \(\frac{1}{3}C_{l_1}^2\) versus the off-diagonal elements of the quantum state.}
	\label{Fig:XY}
\end{figure}

It is clear that \(X\) characterizes the deviation between the wave behavior quantifications from the \(l_p\)-norm to \(l_2\)-norm. It captures the coherence (wave nature) when \(|\rho_{ij}| \neq 0\)  (\(i < j\)) are not all identical, thus serving as a meaningful supplementary quantifier to \(Y\). \(Y\) indicates the quantum coherence distributions. Overall, Eq. \eqref{lp - 4} excels at characterizing strong wave behaviors, while Eq. \eqref{lp-4-1} balances wave, particle and mixedness relationships from the \(l_p\)-norm coherence perspective.
	
\section{Conclusion}\label{sec:section6}

In conclusion, this study has advanced the understanding of wave-particle-mixedness tradeoffs by establishing new \(l_p\)-norm coherence-based relationships. By confirming that the \(l_p\)-norm coherence $(1 \le p\le 2)$ is a valid measure of wave property, we have presented a tradeoff relation among the wave nature (\(l_2\)-norm coherence), mixedness (normalized linear entropy), and particle nature (predictability).  

For general \(l_p\)-norm coherence, we have introduced the quantities $X$ and $Y$ based on the differences between the \(l_p\)-norm and \(l_2\)-norm coherence, and formulated two tradeoff relations satisfied by the particle, wave properties and the mixedness. Taking the \(l_1\)-norm coherence as an example, we have shown that a result given in \cite{Machado2024} is just a special case of our theorem.

Our results highlight the role played by the \(l_p\)-norm coherence in wave-particle-mixedness tradeoffs, clarifying the link from the measure of particle property to the differences between the \(l_2\)-norm maximum coherence and \(l_2\)-norm general coherence, and offering new insights into further investigations on quantum coherence and wave-particle duality.

\section{Ackonwledgments}

This work is being supported by the following: the National Natural Science Foundation of China (NSFC) under Grants 11761073 and 12171044; the Academician Innovation Platform of Hainan Province. We are deeply grateful to Prof. Mingliang Hu for his active participation and invaluable contributions to this research.

\appendix
\section{Proof of Lemma 1}
Here present a simpler alternative proof different from the one given in \cite{Ding2025}. 

As per the work in \cite{Durr2001}, \(V_D(\rho|\Pi)\) is a proper wave measure. Given the relationship \(C_{l_2}(\rho)=\sqrt{\frac{d - 1}{d}}V_D(\rho|\Pi)\), the \(l_2\)-norm coherence is also a legitimate proper wave measure. As a result, we only need prove the case of \(1 \leq p < 2\).

Actually the $l_p$-norm $(1 \le p<2)$ coherence possesses the following favorable properties:

(a) Since \(C_{l_p}(\rho)\) is solely associated with the off-diagonal elements of the quantum state $\rho$, when all the off-diagonal elements are zero, one has \(C_{l_p}(\rho) = 0\), achieving the global minimum.

(b) Suppose \(\rho_{\psi} = \vert\psi\rangle\langle\psi\vert\) is a uniform superposition state, where \(\vert\psi\rangle = \frac{1}{\sqrt{d}}\sum_{i = 1}^{d}\vert i\rangle\). A simple calculation reveals that \(C_{l_p}(\rho_{\psi})=\sqrt[p]{\frac{d(d - 1)}{d^p}}\).

Our goal is to prove that \(C_{l_p}(\rho_{\psi})\) is the global maximum. That is, for any quantum state \(\rho\), the inequality \(C_{l_p}(\rho)\leq\sqrt[p]{\frac{d(d - 1)}{d^p}}\) is satisfied. Consider an arbitrary pure quantum state \(\vert\phi\rangle=\sum_{i = 1}^{d}c_i\vert i\rangle\), to prove \(C_{l_p}(\vert\phi\rangle\langle\phi\vert)\leq C_{l_p}(\rho_{\psi})\) is equivalent to demonstrate that \(\left(\sum_{i\neq j}\vert c_ic_j^*\vert^p\right)^{\frac{1}{p}}\leq\left(\frac{d - 1}{d^{p - 1}}\right)^{\frac{1}{p}}\).

By direct calculation we have
\[
\begin{aligned}
	\sum_{i \neq j} |c_i c_j^*|^p &= \sum_{i,j} |c_i c_j^*|^p - \sum_{i=j} |c_i c_j^*|^p \\
	&= \left(\sum_i |c_i|^p\right)^2 - \sum_i |c_i|^{2p} \\
	&\leq (d - 1) \sum_i |c_i|^{2p} \quad \text{(by Cauchy inequality)} \\
	&\leq \frac{d - 1}{d^{p-1}} \quad \text{(by power mean inequality)},
\end{aligned}
\]
where for the first inequality, we utilized the Cauchy inequality. Namely, for positive vectors \(a\) and \(b\), the Cauchy inequality asserts that \(\left( \sum_i a_i b_i \right)^2 \leq \left( \sum_i a_i^2 \right) \left( \sum_i b_i^2 \right)\). By substituting \(a_i = b_i=|c_i|^{\frac{p}{2}}\) into the formula, we obtain that \(\left(\sum_i |c_i|^p\right)^2 \leq d \sum_i |c_i|^{2p}\).

For the second inequality, we used the power mean inequality. The \(r\)-th power mean of positive real numbers \(a_1,a_2,\dots,a_d\) is defined as \(M_r(a_1,a_2,\dots,a_d)=\left(\frac{1}{d}\sum_{i = 1}^d a_i^r\right)^{\frac{1}{r}}\) for \(r\neq0\). When \(r < s\), \(M_r\leq M_s\), with the equality holding if and only if \(a_1 = a_2=\cdots=a_d\). Given \(\sum_{i=1}^d |c_i|^2 = 1\), we define the power mean 
$$M_r = \left(\frac{\sum_{i=1}^d (|c_i|^2)^r}{d}\right)^{\frac{1}{r}}.$$

For \(1 \le p<2\), the power mean inequality \(M_p \leq M_2\) yields:
\[
\left(\frac{\sum_{i=1}^d |c_i|^{2p}}{d}\right)^{\frac{1}{2p}} \leq \left(\frac{\sum_{i=1}^d |c_i|^2}{d}\right)^{\frac{1}{2}} = \frac{1}{\sqrt{d}},
\]
i.e., 
$$\sum_{i=1}^d |c_i|^{2p} \leq \frac{1}{d^{p-1}}.$$
Thus, \(C_{l_p}(\vert\phi\rangle\langle\phi\vert)\leq C_{l_p}(\rho_{\psi})\).

Now consider any mixed states \(\rho = \sum_k p_k |\psi_k\rangle\langle\psi_k|\) 
with $ |\psi_k\rangle=\sum_{i}c_{k,i}|i\rangle$. The Jensen's inequality gives
\[
\sum_{i \neq j} |\rho_{ij}|^p \leq \sum_k p_k \sum_{i \neq j} |c_{k,i} c_{k,j}^*|^p.
\]
Hence, the \(l_p\)-norm coherence satisfies
\[
C_{l_p}(\rho) = \left(|\rho_{ij}|^p \right)^{\frac{1}{p}} \leq \left(\frac{d - 1}{d^{p-1}}\right)^{\frac{1}{p}} = C_{l_p}(\rho_{\psi}),
\]
confirming that the global maximum of \(C_{l_p}(\rho)\) is attained at the uniform superposition pure state.

(c) Since rearranging the values of $\langle i|\rho|i\rangle$ merely modifies the diagonal elements of the density matrix $\rho$, the $l_{p}$-norm coherence $C_{l_{p}}(\rho)$ remains invariant under such permutations.

(d) For any $\lambda\in[0,1]$ and two quantum states $\rho_1$ and $\rho_2$, the $l_{p}$-norm coherence of the convex combination $\lambda\rho_1+(1 - \lambda)\rho_2$ is given by  
$$C_{l_{p}}(\lambda\rho_1+(1 - \lambda)\rho_2)=\left(\sum_{i\neq j}|\lambda\rho_{1_{ij}}+(1 - \lambda)\rho_{2_{ij}}|^{p}\right)^{\frac{1}{p}}.
$$

The Minkowski inequality, which is valid for $p\geq1$, states that for non-negative real numbers $a_{ij}$ and $b_{ij}$, the following relation holds: 
$$(\sum_{i\neq j}|a_{ij}+b_{ij}|^{p})^{\frac{1}{p}}\leq(\sum_{i\neq j}|a_{ij}|^{p})^{\frac{1}{p}}+(\sum_{i\neq j}|b_{ij}|^{p})^{\frac{1}{p}}.$$ 
By setting $a_{ij}=\lambda\rho_{1_{ij}}$ and $b_{ij}=(1 - \lambda)\rho_{2_{ij}}$, we establish the inequality, $$
C_{l_{p}}(\lambda\rho_1+(1 - \lambda)\rho_2)\leq\lambda C_{l_{p}}(\rho_1)+(1 - \lambda)C_{l_{p}}(\rho_2).
$$ 
Therefore, $C_{l_{p}}$ is convex. Overall, $C_{l_{p}}$ is a well-defined measure of wave property.

\section{Proof of (10)}

We use the H{\"o}lder inequality. For two sequences of non-negative real numbers \(a_i\) and \(b_i\) ($i\leq d$), and positive real numbers \(r, s > 1\) satisfying \(\frac{1}{r} + \frac{1}{s} = 1\), the H{\"o}lder inequality states that
$$
\sum_{i=1}^d a_i b_i \leq \left( \sum_{i=1}^d a_i^r \right)^{\frac{1}{r}} \left( \sum_{i=1}^d b_i^s \right)^{\frac{1}{s}}.
$$
The equality is attained if and only if there exists a constant $k$ such that \(a_i^r = k b_i^s\) for all $i$.

Building on this inequality, for \( 1 \leq p < 2 \) we have
\[
\begin{aligned}
	C_{l_p}^p(\rho)=\sum_{i \neq j}|\rho_{ij}|^p
	&=\sum_{i \neq j}(|\rho_{ij}|^2)^\frac{p}{2}\\
	&\le \left\{\sum_{i \neq j}\left[(|\rho_{ij}|^2)^\frac{p}{2}\right]^\frac{2}{p}\right\}^\frac{p}{2} \left(\sum_{i \neq j}1 \right)^\frac{2-p}{2}\\
	&=C_{l_2}^p(\rho) \cdot [d(d - 1)]^\frac{2 - p}{2},
\end{aligned}
\]
i.e.,
\begin{equation}
	C_{l_{2}}^{2}\left(\rho\right) \ge \left[d\left(d - 1\right)\right]^{\frac{p - 2}{p}}C_{l_{p}}^{2}\left(\rho\right),  (1\leq p<2).
\end{equation}

\section{Proof of (15)}

We first derive the following lower bound of the \(l_1\)-norm coherence in terms of the \(l_2\)-norm coherence,
\begin{equation}\label{l1le}
C_{l_{1}}^{2}(\rho)\geq 2C_{l_{2}}^{2}(\rho).
\end{equation}
Actually,
\begin{align*}
	&  C_{l_{1}}^{2}(\rho)-2C_{l_{2}}^{2}(\rho)\\
	&= \left(\sum_{i\neq j}|\rho_{ij}|\right)^2 - 2\sum_{i\neq j}|\rho_{ij}|^2 \\
	&= \left(\sum_{i\neq j}|\rho_{ij}|\right)\left(|\rho_{ij}| + |\rho_{ji}|+\sum_{\substack {k\neq l \\ (k,l)\neq(i,j) \\ (k,l)\neq (j,i)}}|\rho_{kl}|\right)-2\sum_{i\neq j}|\rho_{ij}|^2 \\
	&= 2\sum_{i\neq j}|\rho_{ij}|^2+\sum_{i\neq j}\sum_{\substack {k\neq l \\ (k,l)\neq(i,j) \\ (k,l)\neq (j,i)}}|\rho_{ij}||\rho_{kl}|-2\sum_{i\neq j}|\rho_{ij}|^2 \\
	&= \sum_{i\neq j}\sum_{\substack {k\neq l \\ (k,l)\neq(i,j) \\ (k,l)\neq (j,i)}}|\rho_{ij}||\rho_{kl}|\geq 0.
\end{align*}

Starting from the definition of $C_{l_{p}}^{2}(\rho)(p\geq 1)$, we have
\begin{align*}
	C_{l_{p}}^{2}(\rho) &= \left[\left(\sum_{i \neq j}|\rho_{ij}|^p\right)^{\frac{1}{p}}\right]^2\\
	&=\left[\left(\sum_{i \neq j}|\rho_{ij}|^p\right)^{\frac{1}{p}}\left(\sum_{i \neq j}1\right)^{\frac{p - 1}{p}}\left(\frac{1}{d(d - 1)}\right)^{\frac{p - 1}{p}} \right]^2.
\end{align*}
By applying the H{\"o}lder inequality, we obtain
\begin{align*}
	C_{l_{p}}^{2}(\rho) &\ge \left[\left(\sum_{i \neq j}|\rho_{ij}|\right)\left(\frac{1}{d(d - 1)}\right)^{\frac{p - 1}{p}} \right]^2\\
	&=\left(\frac{1}{d(d - 1)}\right)^{\frac{2(p - 1)}{p}}\left(\sum_{i \neq j}|\rho_{ij}|\right)^2.
\end{align*}
Then, leveraging the previously derived lower bound \eqref{l1le}, we get:
\begin{equation}
	C_{l_{p}}^{2}(\rho)\ge 2\left(d(d - 1)\right)^{\frac{2(1 - p)}{p}}C_{l_2}^2(\rho).
\end{equation}

\vspace{0.2cm}
\section{References}
\begingroup
\renewcommand{\section}[2]{}%

\endgroup 
\end{normalsize}

\end{document}